\documentclass[aps,prb,twocolumn,showpacs,superscriptaddress,floatfix,10pt]{revtex4-1}

\usepackage{graphicx}
\usepackage{amsmath}
\usepackage{epsfig}
\usepackage{helvet}
\usepackage{amssymb}
\newcommand{\ket}[1]{\mbox{$\big| #1 \big\rangle$}}
\newcommand{\vecb}[1]{\mathord{\buildrel{\lower3pt\hbox{$\scriptscriptstyle\rightarrow$}} \over #1} }
\newcommand{\imp}{\mbox{\tiny imp}}
\newcommand{\Wilson}{\mbox{\tiny W}}

\begin{document}

\title{A theory of minimal updates in holography}

\author{G. Evenbly}
\affiliation{Institute for Quantum Information and Matter, California Institute of Technology, Pasadena CA 91125, USA}
\email{evenbly@caltech.edu}
\author{G. Vidal}
\affiliation{Perimeter Institute for Theoretical Physics, Waterloo, Ontario N2L 2Y5, Canada}  \email{gvidal@perimeterinstitute.ca}
\date{\today}

\begin{abstract}
Consider two quantum critical Hamiltonians $H$ and $\tilde{H}$ on a $d$-dimensional lattice that only differ in some region $\mathcal{R}$. We study the relation between holographic representations, obtained through real-space renormalization, of their corresponding ground states $\ket{\psi}$ and $\ket{\tilde{\psi}}$. We observe that, even though $\ket{\psi}$ and $\ket{\tilde{\psi}}$ disagree significantly both inside and outside region $\mathcal{R}$, they still admit holographic descriptions that only differ inside the past causal cone $\mathcal{C}(\mathcal{R})$ of region $\mathcal{R}$, where $\mathcal{C}(\mathcal{R})$ is obtained by coarse-graining region $\mathcal{R}$. We argue that this result follows from a notion of directed influence in the renormalization group flow that is closely connected to the success of Wilson's numerical renormalization group for impurity problems. At a practical level, directed influence allows us to exploit translation invariance when describing a homogeneous system with e.g. an impurity, in spite of the fact that the Hamiltonian is no longer invariant under translations.
\end{abstract}


\maketitle

\section{Introduction}
The renormalization group (RG) \cite{RGA, RGB, RGC}, fundamental to our conceptual understanding of quantum field theory and critical phenomena, is also the basis of important approaches to many-body problems. In RG methods, the microscopic Hamiltonian of an extended system is simplified through a sequence of coarse-graining transformations until a fixed point of the RG flow is reached. From this scale invariant fixed point, the universal, low energy properties of the phase can then be extracted. The RG is also at the core of certain holographic constructions, where the many-body system is regarded as the boundary of another system in one additional dimension corresponding to scale. A prominent example is the AdS/CFT duality \cite{AdSCFTA, AdSCFTB, AdSCFTC} of string theory, where a conformal field theory (CFT) in $d+1$ space-time dimensions is dual to a gravity theory in anti-de Sitter (AdS) space-time in $d+2$ dimensions.

\begin{figure}[!tbph]
\begin{center}
\includegraphics[width=8.5cm]{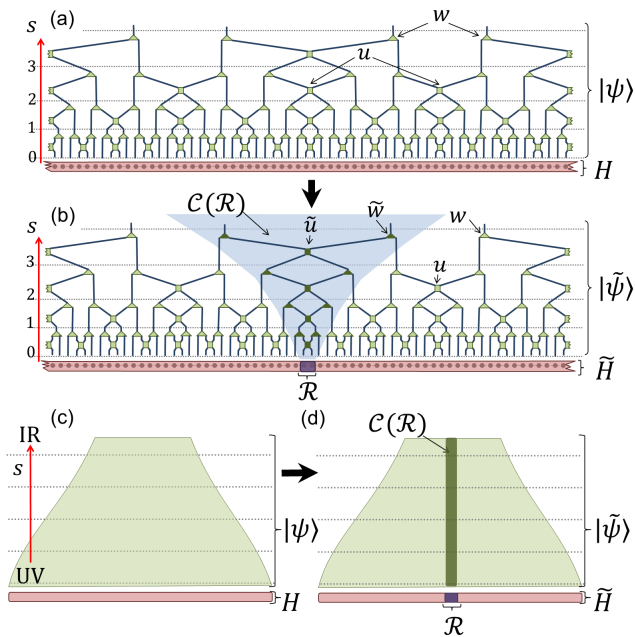}
\caption{ (a) MERA tensor network for the ground state $\ket{\psi}$ of a lattice Hamiltonian $H$ in $d=1$ space dimensions (modified binary scheme of Ref. \onlinecite{LongCFT1}). Scale and translation invariance result in a compact description: two tensors ($u,w$) are repeated throughout the infinite tensor network.
(b) The ground state $\ket{\tilde{\psi}}$ of the Hamiltonian $\tilde{H} = H + H_{\mathcal{R}}^{\imp}$ is represented by a MERA with the same tensors ($u,w$) outside the causal cone $\mathcal C(\mathcal R)$ (shaded). Inside, scale invariance implies again a very compact description: two new tensors ($\tilde{u},\tilde{w}$) repeated throughout the semi-infinite causal cone. (c-d) The same illustrations, without drawing the tensors of the network.}
\label{fig:DirectedMERA}
\end{center}
\end{figure}

Entanglement renormalization \cite{ERA, ERB} is a modern formulation of real-space RG for quantum systems on a lattice, based on the removal of short-range entanglement at each coarse-graining step. By concatenating coarse-graining transformations, one obtains the multi-scale entanglement renormalization ansatz (MERA) \cite{MERA}, an efficient tensor network representation of the many-body ground state, see Fig. \ref{fig:DirectedMERA}(a). The MERA spans an additional dimension corresponding to RG scale and is thus regarded as a lattice realization of holography \cite{WhatIsHolography, Swingle}. Importantly, this tensor network (and generalizations thereof \cite{BranchingA, BranchingB}) is expected to produce a holographic description of \textit{any} many-body system and, in particular, it is not restricted to operate in the so-called \textit{strong coupling, large-$N$} regime that produces a \textit{weakly coupled, semi-classical} gravity dual --- as required in many practical applications of the AdS/CFT correspondence \cite{AdSCFTA, AdSCFTB, AdSCFTC}. As a result, the MERA is a promising tool to gain insights into the structure of holography for a generic many-body system \cite{Swingle, Swingle2, BranchingA, BranchingB, Metric, AllA, AllB, AllC, AllD, AllE, Singh, LongCFT2}, regardless of whether it has e.g. a weakly coupled, semi-classical gravity dual. For instance, in Ref. \onlinecite{LongCFT2} the authors already used the MERA to explore the \textit{modular} character of holography ---namely the possibility of building a holographic description of a complex system by stitching together pieces (or modules) corresponding to the holographic description of simpler systems--- and apply it to the study of critical systems with impurities, boundaries, interfaces, and $Y$-junctions.

In this paper we propose a theory of \textit{minimal updates} in holography. Specifically, we address the following question: Given the ground states $\ket{\psi}$ and $\ket{\tilde{\psi}}$ of two Hamiltonians $H$ and $\tilde{H}$ that only differ in a region $\mathcal{R}$ of a $d$-dimensional lattice $\mathcal{L}$ \cite{dVersusdPlus1}, how much do we have to modify the holographic description of $\ket{\psi}$ in order to produce a holographic description of $\ket{\tilde{\psi}}$? We claim that the answer to this question can be formulated in simple geometric terms: \textit{ A holographic description for $\ket{\tilde{\psi}}$ can be obtained by modifying that of $\ket{\psi}$ only in the causal cone $\mathcal{C}(\mathcal{R})$ of region $\mathcal{R}$}, where $\mathcal{C}(\mathcal{R})$ is the part of the holographic description that traces the evolution of the region $\mathcal{R}$ under coarse-graining. This claim, supported by abundant numerical evidence \cite{LongCFT2}, will be justified here theoretically in terms of a notion of \textit{directed influence} in the RG flow, which we argue to also underpin the success of Wilson's numerical renormalization group (NRG) \cite{RGA, RGB, RGC, NRGreview} for impurity problems. Directed influence leads to an extremely compact, accurate holographic representation of a critical system with an impurity by minimally updating the MERA of a homogeneous system. More generally, as argued in Ref. \onlinecite{LongCFT2}, directed influence implies the modular character of holography.

For concreteness, let us consider a hypercubic lattice $\mathcal{L}$ in $d$ space dimensions, and a particular MERA for the ground states $\ket{\psi}$ and $\ket{\tilde{\psi}}$ of $H$ and $\tilde{H}$ based on a coarse-graining transformation that maps a hyper-cubic block of $2^d$ sites into one effective site, as illustrated for $d=1$ in Fig. \ref{fig:DirectedMERA}(a) \cite{ManyMERASchemes}. We emphasize that the MERA describes both the ground state of the system and a sequence of coarse-graining transformations, where the latter are labeled with a scale parameter $s$, with $s\in \{0,1,2,\cdots\}$. To simplify the notation, we will assume that $H$ is a translation invariant, quantum critical Hamiltonian corresponding to a fixed-point of the RG flow, so that $\ket{\psi}$ is invariant both under translations and changes of scale. Accordingly the MERA for $\ket{\psi}$ can be completely specified by a single pair ($u,w$) of tensors that are repeated throughout the entire tensor network \cite{Giovannetti, Pfeifer, LongCFT1}. [However, the proposed minimal updates do not require translation or scale invariance.]

\section{Causal cones}
The causal cone $\mathcal{C}(\mathcal{R})$ of a region $\mathcal{R}$ of the lattice $\mathcal{L}$, see Fig. \ref{fig:DirectedMERA}(b), was originally defined as the part of the holographic tensor network that can affect the properties of the state $\ket{\psi}$ in region $\mathcal{R}$ \cite{MERA}. The peculiar structure of causal cones in the MERA is the key reason why one can \textit{efficiently} compute expectation values of local observables from this tensor network \cite{Algorithms}. Here we argue that the causal cone $\mathcal{C}(\mathcal{R})$ also defines the region of the MERA that needs to be updated in order to account for a change of the Hamiltonian in region $\mathcal{R}$, Fig. \ref{fig:DirectedMERA}(c)-(d). We emphasize that this new role of the causal cones, of clear physical significance and (as we will argue) ultimately connected to the existence of different energy scales in the Hamiltonian $H$, is unrelated to the computational considerations that guided the design of the MERA \cite{MERA,Algorithms}.
Geometrically, the causal cone $\mathcal{C}(\mathcal{R})$ is the part of the tensor network that contains the evolution of region $\mathcal{R}$ under successive coarse-graining transformations. To further simplify the analysis, we will assume that $\mathcal{R}$ is a hyper-cubic region $\mathcal{R}$ made of $2^d$ sites \cite{ModularityAlsoForLargeR}. This region can be see to be mapped into an identical hyper-cubic region with $2^d$ sites under coarse-graining transformations (see Appendix \ref{ApdxA}).

\section{Minimal update}
Let us now consider the ground state $\ket{\tilde{\psi}}$ of Hamiltonian $\tilde{H} = H + H_{\mathcal{R}}^{\imp}$, where $H_{\mathcal{R}}^{\imp}$ accounts for an impurity on the hyper-cubic region $\mathcal{R}$ made of $2^d$ sites. Our claim is that a MERA for $\ket{\tilde{\psi}}$ can be obtained from the MERA for the ground state $\ket{\psi}$ in the absence of the impurity by simply replacing, inside the causal cone $\mathcal{C}(\mathcal{R})$, the tensors ($u$, $w$) with new tensors. Specifically, if the impurity is itself already a new RG fixed-point (e.g. a conformal defect in a CFT \cite{ConformalDefect}), which implies that $\ket{\tilde{\psi}}$ is still scale invariant, then the entire causal cone $\mathcal{C}(\mathcal{R})$ can be completely specified by a single new pair ($\tilde{u}$, $\tilde{w}$) of tensors, see Fig. \ref{fig:DirectedMERA}(b).

In Ref. \onlinecite{LongCFT2,BoundaryMERA} we have presented abundant numerical evidence supporting the validity of the proposed minimal update, and have argued that this construction naturally reproduces: ($i$) the power-law scaling of expectation values of local observables (e.g. of the local magnetization in the case of a magnetic impurity) with the distance to the impurity; and ($ii$) the set of new scaling operators and scaling dimensions attached to the impurity \cite{ConformalDefect}. The above compact description in terms of just two pairs of tensors $\{(u, w), (\tilde{u}, \tilde{w})\}$, valid even in the thermodynamic limit, is somewhat surprising. After all, one would expect that coarse-graining the impurity system, which is not translation invariant, would require the use of different coarse-graining tensors $(u(x), w(x))$ at different locations $x$ of lattice $\mathcal{L}$. Accordingly, the number of variational parameters, proportional to the number of different tensors in the MERA, would grow linearly in the size of the system. Instead, by only updating the causal cone $\mathcal{C}(\mathcal{R})$ [which amounts to exploiting the translation invariance of Hamiltonian $H$ to describe the ground state of $\tilde{H}$] we can address an impurity system directly in the thermodynamic limit, and thus avoid finite size effects when extracting the universal properties of the critical impurity \cite{BoundaryMERA,LongCFT2}. 

\begin{figure}[!tbph]
\begin{center}
\includegraphics[width=8.5cm]{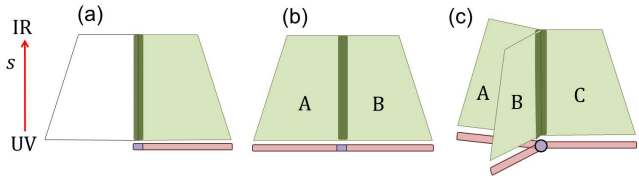}
\caption{
Updating only the causal cone of $\mathcal{R}$ also produces a simple holographic description of a scale-invariant boundary, in terms of tensors $(u,w)$ in the bulk and a boundary tensor $\tilde{w}$, as described in Ref. \onlinecite{BoundaryMERA}. More complex systems, such as (b) an interface, and (c) a $Y$-junction, can be similarly described by a \textit{modular} MERA, consisting of a bulk tensors $(u_{\alpha}, w_{\alpha})$ for each type of material $\alpha$ ($\alpha = A,B,\cdots$) and defect tensors $(\tilde{u},\tilde{w})$ that glue the different modules together \cite{LongCFT2}.
}
\label{fig:Defects}
\end{center}
\end{figure}

Similar constructions are also possible for more complex systems, including systems with a boundary, an interface or a $Y$-junction, see Fig. \ref{fig:Defects}, for which recursive application of minimal updates leads to the modular MERA, as discussed in \cite{LongCFT2}. Below we shall argue that the validity of the proposed minimal update follows a more fundamental property of RG flows, that we call directed influence. In order to discuss the latter, we must first introduce an effective lattice model that describes the causal cone $\mathcal{C}(\mathcal{R})$.

\section{Wilson chain}
\begin{figure}[!tbph]
\begin{center}
\includegraphics[width=8.5cm]{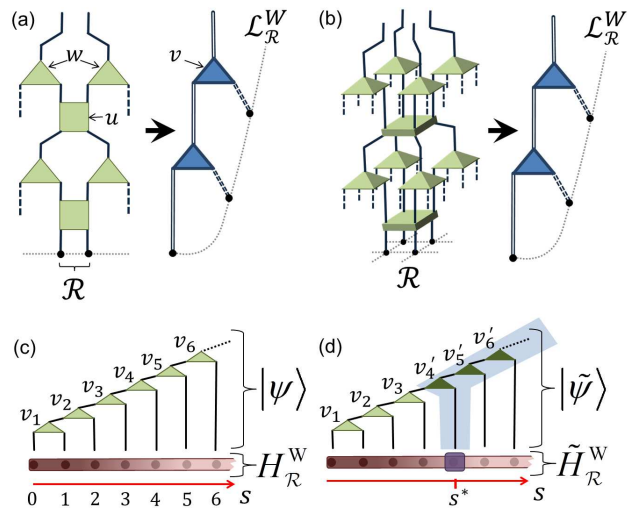}
\caption{
(a) Tensors inside the causal cone $\mathcal{C}(\mathcal{R})$ in $d=1$ dimensions. Site $s$ of the Wilson chain $\mathcal{L}^{\Wilson}_{\mathcal{R}}$ corresponds to the two effective sites at scale $s$. By replacing three tensors $(u,w,w)$ with a single tensor $v$, we obtain an MPS representation of the ground state of $H^{\Wilson}_{\mathcal{R}}$.
(b) Equivalent construction in $d=2$ dimensions. In this case, 12 effective sites at scale $s$ become a single site of $\mathcal{L}^{\Wilson}_{\mathcal{R}}$, whereas each MPS tensor $v$ corresponds to five tensors $(u,w,w,w,w)$.
(c)-(d) Directed influence: changing the Wilson chain Hamiltonian $H^{\Wilson}_{\mathcal{R}}$ on site $s^{*}$ results in a new ground state MPS where only tensors $v_s$ for scales $s\geq s^*$ are updated.
}
\label{fig:DirectedMPS}
\end{center}
\end{figure}

We call the \textit{Wilson chain} of region $\mathcal{R}$, denoted $\mathcal{L}^{\Wilson}_{\mathcal{R}}$, the semi-infinite, one-dimensional lattice built by coarse-graining the $d$-dimensional lattice $\mathcal{L}$ by all the tensors in the MERA that lay outside the causal cone $\mathcal{C}(\mathcal{R})$. More precisely, each site of the Wilson chain $\mathcal{L}^{\Wilson}_{\mathcal{R}}$ is uniquely labeled by a value of the scale parameter $s$ and it collects together all the effective sites at scale $s$ obtained through the above coarse-graining, see Fig. \ref{fig:DirectedMPS}(a)-(b). By construction, site $s$ of $\mathcal{L}^{\Wilson}_{\mathcal{R}}$ effectively represents the $O(2^{ds})$ sites of $\mathcal{L}$ located roughly at a distance [measured in lattice spacing] $2^s$ away from region $\mathcal{R}$. Thus, progressing from site $s$ to site $s+1$ of the Wilson chain $\mathcal{L}^{\Wilson}_{\mathcal{R}}$ corresponds to simultaneously increasing length scale and moving away from region $\mathcal{R}$.

The Wilson chain is equipped with an effective Hamiltonian $H^{\Wilson}_{\mathcal{R}}$, obtained by coarse-graining $H$, of the form
\begin{equation}\label{eq:hWilson}
    H^{\Wilson}_{\mathcal{R}} = h_{[0]} + \sum_{s=0}^{\infty} \Lambda^{-s} h^{\Wilson}_{[s,s+1]}.
\end{equation}
The nearest neighbor term $\Lambda^{-s} h^{\Wilson}_{[s,s+1]}$ consists of a two-site hermitian operator $h^{\Wilson}_{[s,s+1]}$ that is \textit{independent} of $s$, multiplied by a negative power of an amplitude $\Lambda > 1$, which takes the value $\Lambda=2^{z}$, where $z$ is the dynamic critical exponent of $H$ (e.g. $z=1$ for Lorentz invariant quantum critical points), see Appendix \ref{ApdxD} for details and also Refs. \onlinecite{LongCFT2,BoundaryMERA} for complimentary derivations of the effective Hamiltonian for the Wilson chain.

The structure of the one-dimensional Hamiltonian $H^{\Wilson}_{\mathcal{R}}$, with exponentially decaying nearest-neighbor terms, is similar to that obtained by Wilson as part of his resolution of the Kondo impurity problem -- a single impurity in a three dimensional bath of three fermions \cite{RGA, RGB, RGC}. However, we note that while having a free fermion bath was key in Wilson's derivation of an effective one-dimensional lattice model, here we use the MERA to (at least in principle) address non-perturbatively any type of $d$-dimensional bath.

\section{Directed influence} Following Wilson's NRG method \cite{RGA, RGB, RGC, NRGreview} (see Appendix \ref{ApdxB}), the ground state of $H^{\Wilson}_{\mathcal{R}}$ can be obtained by identifying, progressing iteratively over $s$, the low energy subspace $\mathbb{H}_s$ of the first $s+1$ sites of $\mathcal{L}^{\Wilson}_{\mathcal{R}}$,
\begin{equation}\label{eq:Hsub}
    \mathbb{H}_s \subseteq \mathbb{V}_0 \otimes \mathbb{V}_1 \otimes \cdots \otimes \mathbb{V}_s,
\end{equation}
where $\mathbb{V}_s$ is the vector space of site $s$ in $\mathcal{L}^{\Wilson}_{\mathcal{R}}$. More specifically, $\mathbb{H}_s$ is chosen (by means of a suitable energy minimization) to be the low energy subspace of $\mathbb{H}_{s-1} \otimes \mathbb{V}_{s}$, and is characterized by a linear map $v_s$,
\begin{equation}\label{eq:v}
    v_s:\mathbb{H}_s \rightarrow \mathbb{H}_{s-1} \otimes \mathbb{V}_s.
\end{equation}
Then the tensors $\{v_1,v_2, \cdots\}$ form a \textit{matrix product state} (MPS) \cite{MPSA, MPSB} representation of the ground state of $H^{\Wilson}_{\mathcal{R}}$.

For the present purposes, the most important feature of the NRG method is that the low energy subspace $\mathbb{H}_s$ (equivalently, tensor $v_s$) only depends on the restriction of the Hamiltonian to sites $\{0,1, \cdots, s\}$,
\begin{equation}\label{eq:terms}
    h_{[0]} + \Lambda^0 h^{\Wilson}_{[0,1]} + \Lambda^{-1} h^{\Wilson}_{[1,2]} + \cdots + \Lambda^{-s+1} h^{\Wilson}_{[s-1,s]},
\end{equation}
and not on the Hamiltonian terms related to larger length scales. In other words, if we modify the Hamiltonian at some site $s^*$, then NRG produces an MPS representation of the new ground state where only the tensors $v_s$ for $s\geq s^*$ are modified, see Fig. \ref{fig:DirectedMPS}(c)-(d). That is, assuming the validity of the NRG approach, changes in the Hamiltonian at length scale $s^{*}$ only affect the ground state representation at larger length scales, a property that we refer to as \textit{directed influence} in the RG flow. We emphasize that the validity of Wilson's NRG, and thus also directed influence, relies heavily on the factor $\Lambda^{-s}$ to induce a separation of energy scales in the problem. When such a separation of energy scales is present, then the treatment of one energy scale at a time as prescribed by the NRG approach can be justified from perturbation theory (see Appendix \ref{ApdxB}). In the absence of such a factor the NRG approach would typically fail \cite{DMRG}, such that directed influence would also fail, and a change in the Hamiltonian at length scale $s^{*}$ could affect the ground state representation at all length scales $s$.

We are finally ready to show that the validity of the proposed minimal update in the MERA follows from assuming the validity of directed influence in Wilson chains. Let us modify the Hamiltonian from $H$ to $\tilde{H} = H + H^{\imp}_{\mathcal{R}}$ to account for an impurity, and study how the ground state MPS for different Wilson chains (corresponding to different regions of lattice $\mathcal{L}$) must react to this change according to directed influence.

\begin{figure}[!tbph]
\begin{center}
\includegraphics[width=8.5cm]{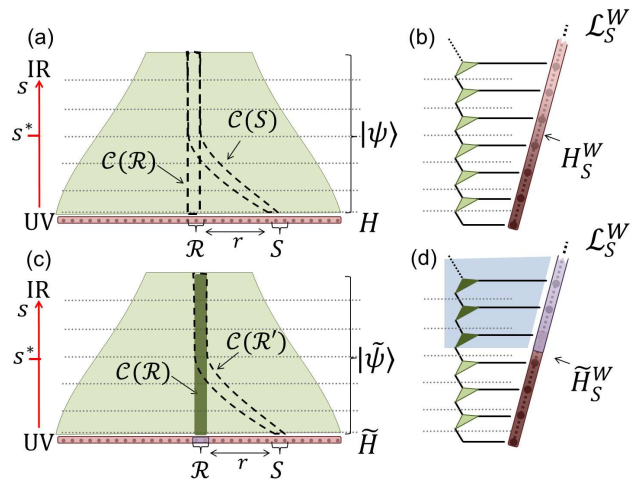}
\caption{
(a) The causal cones $\mathcal{C}(\mathcal{R})$ and $\mathcal{C}(\mathcal{S})$ for two regions $\mathcal{R}$ and $\mathcal{S}$ separated by $r$ sites become coincident at scale $s^* \approx \log_2(r)$.
(b) MPS representation of the ground state in the Wilson chain $\mathcal{L}^{\Wilson}_{\mathcal{S}}$.
(c) In the presence of an impurity in region $\mathcal{R}$, a minimal update change only the tensors inside $\mathcal{C}(\mathcal{R})$. This amounts to changing the tensors inside $\mathcal{C}(\mathcal{S})$ only at scales $s\geq s^{*}$.
(d) Directed influence justifies this minimal update of tensors in the MERA: Hamiltonians $H^{\Wilson}_{\mathcal{S}}$ and $\tilde{H}^{\Wilson}_{\mathcal{S}}$ only differ at scale $s^{*}$, and therefore the tensors in the causal cone $\mathcal{C}(\mathcal{S})$ indeed only need to be updated at length scales $s\geq s^{*}$.
}
\label{fig:MERAproof}
\end{center}
\end{figure}

First, we notice that the effective Hamiltonian of the Wilson chain $\mathcal{L}^{\Wilson}_{\mathcal{R}}$ for the causal cone $\mathcal{C}(\mathcal{R})$ is modified from $H^{\Wilson}_{\mathcal{R}}$ in Eq. \ref{eq:hWilson} to
\begin{equation}\label{eq:htildeWilson}
    \tilde{H}^{\Wilson}_{\mathcal{R}} = \tilde{h}_{[0]} + \sum_{s=0}^{\infty} \Lambda^{-s} h^{\Wilson}_{[s,s+1]}.
\end{equation}
where $\tilde{h}_{[0]} = h_{[0]} + H^{\imp}_{\mathcal{R}}$ includes the impurity Hamiltonian. In this case, directed influence tells us to change all the tensors in the MPS representation of the ground state of $\tilde{H}^{\Wilson}_{\mathcal{R}}$. This is equivalent to the announced modification of all the tensors in the causal cone $\mathcal{C}(\mathcal{R})$ of the MERA representation of the ground state $\ket{\tilde{\psi}}$ of $\tilde{H}$.

Second, let us consider the causal cone $\mathcal{C}(\mathcal{S})$ of another small region $\mathcal{S}$ of the original lattice $\mathcal{L}$ that is diplaced $\vec r$ from $\mathcal R$, and the corresponding Wilson chain $\mathcal{L}^{\Wilson}_{\mathcal{S}}$, see Fig. \ref{fig:MERAproof}. Here we see that the effective Hamiltonians $H^{\Wilson}_{\mathcal{S}}$ and $\tilde{H}^{\Wilson}_{\mathcal{S}}$, corresponding to the homogeneous and impurity systems respectively, only differ at the length scale $s^* \sim \log_2 |\vec r|$, where the causal cones $\mathcal{C}(\mathcal{R})$ and $\mathcal{C}(\mathcal{S})$ become coincident. Directed influence implies that the MPS representations of the ground states of $H^{\Wilson}_{\mathcal{S}}$ and $\tilde{H}^{\Wilson}_{\mathcal{S}}$ only need to differ at scales $s\geq s^*$, i.e. that the tensors in $\mathcal C(\mathcal S)$ at scales $s<s^*$, which lie outside of the causal cone $\mathcal C(\mathcal R)$, may be left unchanged. This argument is general for any local region $\mathcal S$, thus justifying the proposal of minimal updates in MERA.

\section{Discussion} The holographic description of a many-body system based on real-space RG is not unique. Since the MERA is built by concatenating several coarse-graining transformations, there is indeed some freedom as to how we choose to coarse-grain the system at a given length scale, provided that we compensate for our choice when coarse-graining the system at larger length scales. The minimal update discussed in this paper corresponds to a particular choice of this freedom in coarse-graining. By restricting the update to the causal cone $\mathcal{C}(\mathcal{R})$ of region $\mathcal{R}$, an impurity that is initially localized in space remains localized in space under coarse-graining, and this leads to a very efficient holographic description of $\ket{\tilde{\psi}}$ \cite{NoDMRG}. We conclude with remarks on how the structure of minimal updates in the MERA may translate into a property of the AdS/CFT correspondence \cite{AdSCFTA, AdSCFTB, AdSCFTC}. It is natural to speculate that the non-uniqueness of MERA descriptions, originating in the freedom existing in real-space coarse-graining, is closely related to diffeomorphism invariance in the bulk of the gravity dual. Accordingly, the minimal updates discussed in this paper would be possible also in the gravity dual of the AdS/CFT correspondence after a proper choice of gauge. However, making these ideas more concrete may first require a better understanding of the bulk metric in the MERA \cite{Metric,Swingle2}.

The authors thank Davide Gaiotto, Rob Myers, and Brian Swingle for insightful comments. G.E. is supported by the Sherman Fairchild Foundation.

\newpage

\appendix
\section{Causal cones in the MERA} \label{ApdxA}
In this appendix we describe the structure of causal cones in the MERA. The causal cone $\mathcal C(\mathcal R)$ of a local region $\mathcal R$ is the part of the tensor network that contains the evolution of the region under successive coarse-graining transformations. Causal cones in MERA have a characteristic form, resulting from the peculiar structure of the tensor network, as we now examine.

\begin{figure}[!tbph]
\begin{center}
\includegraphics[width=8cm]{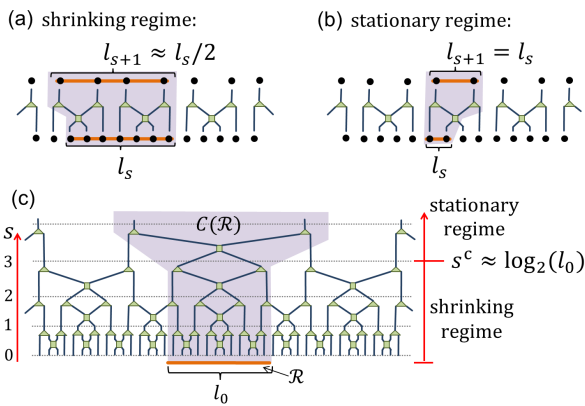}
\caption{(a) A region of $l_s \gg 1$ sites is coarse-grained under a layer of MERA to a smaller region of $l_{s+1} \approx l_s /2$ sites. (b) A region of $l_s =2 $ sites is coarse-grained under a layer of MERA to a region of equivalent width, i.e. $l_{s+1} = l_s = 2$. (c) The width of the causal cone $\mathcal C(\mathcal R)$ of a region $\mathcal R$ comprised of $l_0 \gg 1$ sites shrinks with increasing scale $s$ until the crossover scale $s^{\textrm c}\approx \log_2(l_0)$ is reached, after which it remains stationary.}
\label{fig:CausalRegime}
\end{center}
\end{figure}

We consider the specific MERA scheme analyzed in the main text, namely the \textit{modified binary} MERA on a one-diemsnional lattice $\mathcal L$ \cite{ManyMERASchemes}. Let $\mathcal R$ be a region of $l_0$ contiguous sites in $\mathcal L$, and let $l_s$ be the number of effective sites contained within the causal cone $\mathcal C(\mathcal R)$ at depth $s$. In a single step of the coarse-graining transformation, the disentanglers $u$ act to spread the support of the causal cone by at most two sites, while the isometries $w$ act to compress the support by roughly a factor of two. If $l_s \gg 1$ sites are enclosed by the causal cone at depth $s$ then, under a layer of coarse graining, the action of the isometries dominates and the support of the causal cone shrinks by roughly a factor of two, i.e. $l_{s+1} \approx l_s /2$, see Fig. \ref{fig:CausalRegime}(a). We refer to this as the \emph{shrinking regime} of the causal cone. Conversely, if $l_s = 2$ then the spread of the support from the disentanglers is exactly balanced by the shrinking of the support from the isometries, and the causal cone remains at a fixed width, i.e. $l_{s+1} = l_s $. We refer to this as the \emph{stationary regime} of the causal cone. Thus the causal cone $\mathcal C(\mathcal R)$ of a region $\mathcal R \in \mathcal L$ of $l_0 \gg 1$ sites is in the shrinking regime up to some crossover depth $s^{\textrm c} \approx \log_2 (l_0)$ after which it remains in the stationary regime, see Fig. \ref{fig:CausalRegime}(c).

\section{The numerical renormalization group} \label{ApdxB}
In this appendix we review Wilson's numerical renormalization group (NRG) \cite{RGA, RGB, RGC,  NRGreview}. First we recount Wilson's original arguments justifying the validity of the approach. Then we describe the technical details of its implementation. 

NRG is a method for computing the low energy subspace of a one-dimensional lattice Hamiltonian of the form,
\begin{equation}
    H^{\Wilson} = h_{[0]} + \sum_{s=1}^{s_\textrm{max}} \Lambda^{-s} h^{\Wilson}_{[s-1,s]}, \label{eq:A1}
\end{equation}
that we shall refer to as a Wilson chain Hamiltonian. The nearest neighbor term $\Lambda^{-s} h^{\Wilson}_{[s,s+1]}$ consists of a two site Hermitian operator $h^{\Wilson}_{[s,s+1]}$ that is independent of $s$ multiplied by the negative power of an amplitude $\Lambda>1$. Note that this is the form of the the effective Hamiltonian obtained, and subsequently solved, by Wilson in his solution to the Kondo impurity problem. For concreteness we shall henceforth set $\Lambda=10$ (although the following arguments remain valid for any $\Lambda>1$) and also assume that each site $s$ in $\mathcal L$ is associated to a two dimensional vector space $\mathbb V_s$, such that $h^{\Wilson}_{[s,s+1]}$ could be represented as a $4\times 4$ hermitian matrix. We further assume that the spacing of the eigenvalues of $h^{\Wilson}$ is of order unity.

\begin{figure}[!tbph]
\begin{center}
\includegraphics[width=8cm]{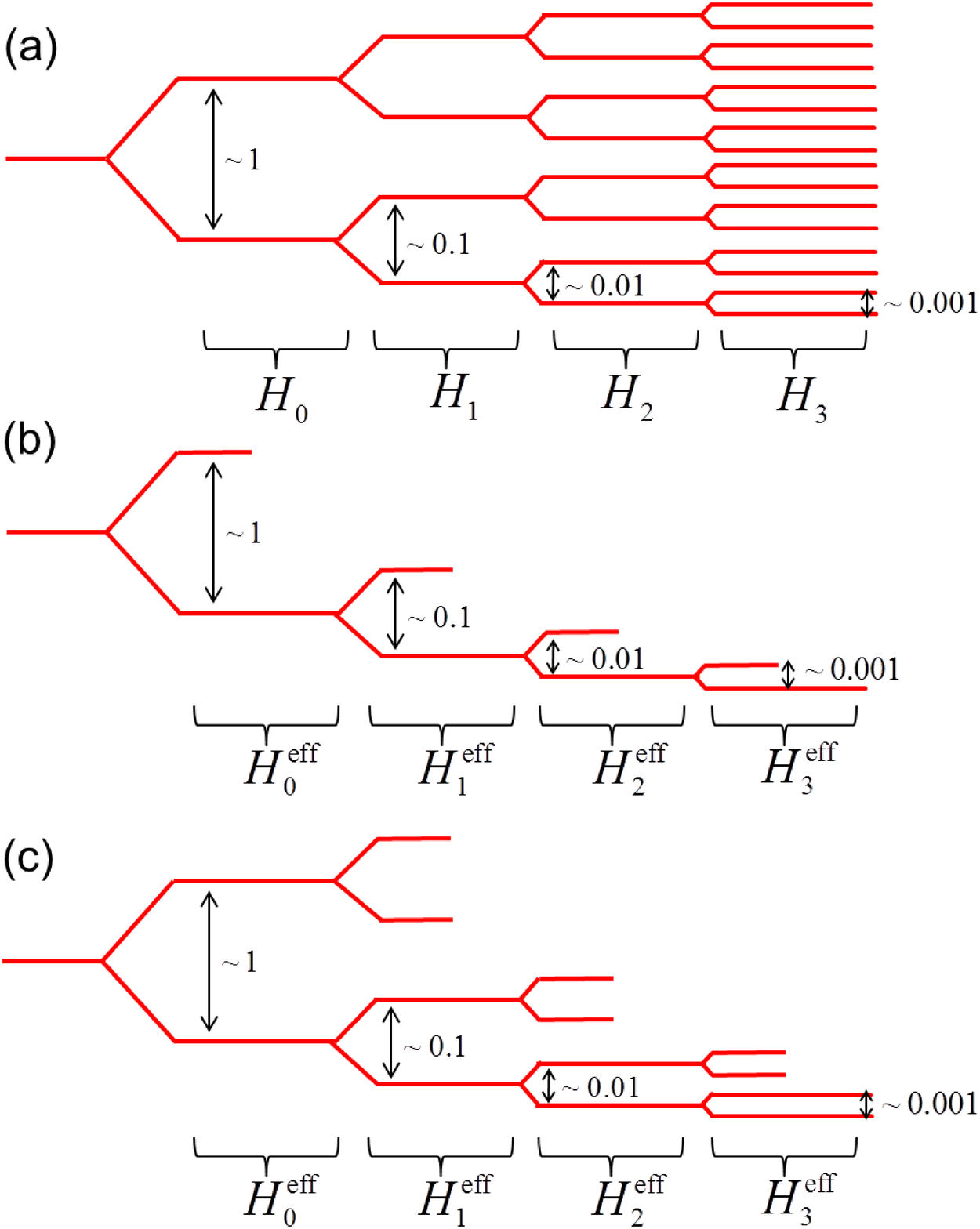}
\caption{(a) Approximate energy levels, from a perturbative analysis, of the sequence of block Hamiltonians $H_s$ (as defined Eq. \ref{eq:A2}), each of which corresponds to the part of the Wilson chain $H^{\Wilson}$ supported on the first $s+1$ sites of the lattice. (b) Approximate energy levels of the sequence of effective Hamiltonians $H_s^\textrm{eff}$ generated by NRG (fixing $\chi=1$), where each $H_s^\textrm{eff}$ is obtained by projecting $H_s$ onto a two dimensional subspace. (c) Approximate energy levels of the sequence of effective Hamiltonians $H_s^\textrm{eff}$ generated by NRG (fixing $\chi=2$), where each $H_s^\textrm{eff}$ is obtained by projecting $H_s$ onto a four dimensional subspace.}
\label{fig:EnergySplit}
\end{center}
\end{figure}

It is possible to understand the qualitative features of the energy spectrum of a Wilson chain Hamiltonian just from the peculiar form the Hamiltonian takes (without the need to specify the local interactions $h^{\Wilson}$), as we now discuss. First we define, for $s\le s_\textrm{max}$, the block Hamiltonian $H_s$ as the part of the original Hamiltonian $H^{\Wilson}$ that is supported on sites $[0,1,\cdots,s]$, with Hilbert space
\begin{equation}
{\mathbb{V}_0} \otimes {\mathbb{V}_1} \otimes {\mathbb{V}_2} \otimes  \cdots  \otimes {\mathbb{V}_s},
\end{equation}
and which consists of terms
\begin{equation}
H_{s} \equiv {h_{[0]}} + \frac{1}{10} h^{\Wilson}_{[0,1]} + \frac{1}{10^2} h^{\Wilson}_{[1,2]} +\ldots +\frac{1}{10^s} h^{\Wilson}_{[s-1,s]}. \label{eq:A2}
\end{equation}
Notice that the final block Hamiltonian in the series reproduces the full Wilson chain Hamiltonian, i.e. ${H_{{s_{\max }}}} = H^{\Wilson}$. Perturbation theory can be used to gain an understanding of the Wilson chain by treating the local couplings in $H^{\Wilson}$ with small prefactors (i.e. those at greater distance $s$ from the start of the chain) as perturbations of the local couplings with larger prefactors, as we now describe. By assumption the first block Hamiltonian, $H_0 \equiv h_{[0]}$, has two energy levels that differ in magnitude by order unity. The spectrum of $H_1=H_0 + \frac{1}{10} h_{[0,1]}$ can be understood by considering $\frac{1}{10}h_{[0,1]}$ as a perturbation of $H_0$; the two energy levels of $H_0$ are each split into two further levels that differ by $\sim \tfrac{1}{10}$. Likewise, we could then understand the spectrum of $H_2=H_1 + \frac{1}{10^2} h_{[1,2]}$ by considering $\frac{1}{10}h_{[1,2]}$ as a perturbation of $H_1$ that splits each of its energy levels by $\sim \tfrac{1}{100}$ etc. Thus the spectrum of $H_s$ for any $s$, and, by extension the full Hamiltonian $H^{\Wilson}$, generically takes the form as shown Fig. \ref{fig:EnergySplit}(a). The numerical renormalization group (NRG) formalizes this perturbative understanding of Wilson chains into an algorithm for their solution.

\begin{figure}[!tbph]
\begin{center}
\includegraphics[width=8.5cm]{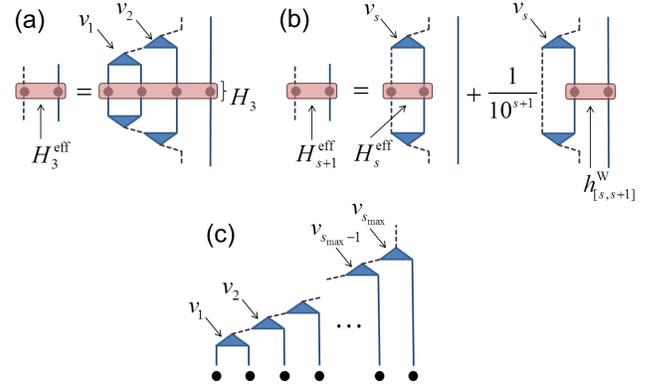}
\caption{(a) An effective Hamiltonian $H^\textrm{eff}_3$ for the block Hamiltonian $H_3$ is obtained by projecting it onto the subspace $\left( \mathbb H_3 \otimes \mathbb V_3 \right) \subseteq \left( \mathbb V_0 \otimes \mathbb V_1 \otimes \mathbb V_2 \otimes \mathbb V_3 \right)$ with isometries $v_1$ and $v_2$. (b) The effective Hamiltonian $H^\textrm{eff}_{s+1}$ can be obtained from $H^\textrm{eff}_{s}$ through addition of the local coupling $h_{[s,s+1]}^{\Wilson}$ and subsequent application of the isometry $v_s$ (obtained through diagonalization of $H^\textrm{eff}_{s}$). (c) The mapping from the original Hilbert space to the low energy subspace of the Wilson chain is a product of isometries $v_s$, which one can recognize as a matrix product state (MPS) of bond dimension $\chi$.}
\label{fig:NRGadvanced}
\end{center}
\end{figure}

Consider that we are interested in identifying a $\chi$ dimensional subspace $\mathbb H_{s_\textrm{max}}$ of the Hilbert space of $\mathcal L$ (where $\chi$ is some adjustable refinement parameter),
\begin{equation}
{\mathbb{H}_{{s_{\max }}}} \subseteq {\mathbb{V}_0} \otimes {\mathbb{V}_1} \otimes  \cdots  \otimes {\mathbb{V}_{{s_{\max }}}},
\end{equation}
such that the Wilson chain $H^{\Wilson}$, when projected onto this subspace, is an effective Hamiltonian that retains the proper low energy physics of the original. The NRG algorithm allows one to identify such a subspace through a sequence of steps; initially a subspace of first two lattice sites is identified,
\begin{equation}
{\mathbb{H}_1} \subseteq {\mathbb{V}_0} \otimes {\mathbb{V}_1},
\end{equation}
and then, sequentially for all $s\le s_\textrm{max}$, subspaces of larger lattice regions are identified,
\begin{equation}
{\mathbb{H}_s} \subseteq {\mathbb{H}_{s - 1}} \otimes {\mathbb{V}_s},
\end{equation}
where each subspace ${\mathbb{H}_s}$ is restrained to be (at most) $\chi$-dimensional.

The NRG algorithm prescribes that each subspace ${\mathbb{H}_s}$ can be chosen through consideration of only the part of the Hamiltonian $H_s$ that is supported on this block, ignoring the Hamiltonian terms from outside the block. In the first step the subspace ${\mathbb{H}_1} \subseteq \mathbb V_0 \otimes \mathbb V_1$ is chosen by diagonalizing the block Hamiltonian $H_1= h_{[0]}+\tfrac{1}{10} h_{[0,1]}^{\Wilson}$ and retaining the space spanned by its (at most) $\chi$ eigenvectors of lowest energy. Then an isometry $v_1$ is formed from these eigenvectors which serves as a mapping to the reduced Hilbert space,
\begin{equation}
{v_1}:{\mathbb{H}_1} \to {\mathbb{V}_0} \otimes {\mathbb{V}_1}.
\end{equation}
We now use isometry $v_1$ to obtain an effective block Hamiltonian $H_2^\textrm{eff} = v_1^{\dagger} H_2 v_1$ for the initial block Hamiltonian $H_2$ of the first three lattice sites. Notice that, whereas the initial block Hamiltonian $H_2$ is defined on the Hilbert space ${\mathbb{V}_0} \otimes {\mathbb{V}_1} \otimes {\mathbb{V}_2}$, the effective Hamiltonian $H_2^\textrm{eff}$ is defined on the subspace ${\mathbb{H}_{1}} \otimes {\mathbb{V}_2}$.

This process is then iterated over larger blocks; one would next identify the subspace ${\mathbb{H}_2} \subseteq \mathbb H_1 \otimes \mathbb V_2$ by forming an isometry $v_2$ from the span of the $\chi$ lowest energy eigenvectors of $H_2^\textrm{eff}$,
\begin{equation}
{v_2}:{\mathbb{H}_2} \to {\mathbb{H}_1} \otimes {\mathbb{V}_2}.
\end{equation}
The isometry $v_2$ can then be used to generate an effective block Hamiltonian $H_3^\textrm{eff}= v_2^\dagger v_1^\dagger H_3 v_1 v_2$ from the original block Hamiltonian $H_3$, see Fig. \ref{fig:NRGadvanced}(a). Alternatively, the effective block Hamiltonian $H_3^\textrm{eff}$ can be obtained from the previous block Hamiltonian $H_2^\textrm{eff}$ as,
\begin{equation}
  H_{3}^{{\text{eff}}} \equiv v_2^{\dag} \left( H_2^\textrm{eff} \otimes \mathbb{I}_{3} \right) v_2   +   v_2^\dag  \left( \mathbb{I}_{1} \otimes \tfrac{1}{10^2} h_{[2,3]}^{\Wilson} \right) v_2 ,
\end{equation}
see also Fig. \ref{fig:NRGadvanced}(b).

Likewise in subsequent steps, for all $s\le s_\textrm{max}$, each effective Hamiltonian $H^\textrm{eff}_s$ (which equates to the block Hamiltonian $H_s$ projected onto the subspace ${\mathbb{H}_{s - 1}} \otimes {\mathbb{V}_s}$) is diagonalized and an isometry $v_s$ is formed from its $\chi$ eigenvectors of lowest energy. The isometry projects to the subspace ${\mathbb{H}_s}$,
\begin{equation}
{v_s}:{\mathbb{H}_s} \to {\mathbb{H}_{s - 1}} \otimes {\mathbb{V}_s},
\end{equation}
and is used to generate next effective Hamiltonian $H^\textrm{eff}_{s+1}$ as,
\begin{equation}
  H_{s+1}^{{\text{eff}}} \equiv   v_s^{\dag} \left(  H_s^\textrm{eff} \otimes \mathbb{I}_{s+1} \right) v_s  
  + v_s^\dag  \left( \mathbb{I}_{s-1} \otimes \tfrac{1}{10^{s+1}} h_{[s,s+1]}^{\Wilson} \right) v_s ,\label{eq:A11}
\end{equation}
see again Fig. \ref{fig:NRGadvanced}(b), where $\mathbb{I}_{s-1}$ and $\mathbb{I}_{s+1}$ here denote the identity on Hilbert spaces ${\mathbb{H}_{s - 1}}$ and ${\mathbb{V}_{s + 1}}$ respectively. Thus the NRG algorithm generates a sequence of isometric tensors $v_s$ each, in general, mapping from a Hilbert space of dimension $2\chi$ to one of dimension $\chi$, whose product identifies the low energy subspace ${\mathbb{H}_{{s_{\max }}}}$ of the Wilson chain,
\begin{equation}
v_0\cdot v_1 \cdot \ldots \cdot v_{s_{\max }} :{\mathbb{H}_{{s_{\max }}}} \to {\mathbb{V}_0} \otimes {\mathbb{V}_1} \otimes  \cdots  \otimes {\mathbb{V}_{{s_{\max }}}}.
\end{equation} 
Notice that this sequence of isometries $v_s$ form a matrix product state (MPS) of bond dimension $\chi$, see Fig. \ref{fig:NRGadvanced}(c).

A key aspect of the NRG algorithm is that the low energy subspace ${\mathbb{H}_s}$ of the block of the first $s$ lattice sites is chosen only through consideration of the part of the Hamiltonian $H_s$ that is supported on this block (while ignoring the Hamiltonian terms outside of the block). As discussed in the main text, this leads to a notion of directed influence in Wilson chains, which justifies the proposal of minimal updates.

The validity of the NRG algorithm is justified from perturbation theory: in identifying the low energy subspace of a block (consisting of the first $s$ sites of the Wilson chain of Eq. \ref{eq:A1}) only the part of the Hamiltonian within the block need be considered as all couplings that are outside of the block are weaker a factor of $\Lambda^{-1}$ (where it was assumed $\Lambda>1$). In the limit that the perturbation parameter $\Lambda$ approaches unity the method becomes less effective, and typically a subspace $\mathbb H_s$ of large local dimension $\chi$ must be retained in order to maintain accuracy. In his solution to the Kondo impurity problem \cite{RGA, RGB, RGC}, where the energy scale parameter of the Wilson chain was $\Lambda=\sqrt{2}$, Wilson retained $\chi>1000$ states in each effective Hamiltonian in order to achieve an accuracy of a few percent. In contemporary applications of NRG \cite{NRGreview} it is computationally feasible to take $\chi$ at least an order of magnitude larger. If one has $\Lambda=1$ exactly, such that all local couplings are of the same magnitude (as in the case of a homogeneous chain), then the NRG approach is no longer justified and would likely fail \cite{DMRG}.

\section{Scale invariance in the modified binary MERA} \label{ApdxC}
\begin{figure}[!tbph]
\begin{center}
\includegraphics[width=8.5cm]{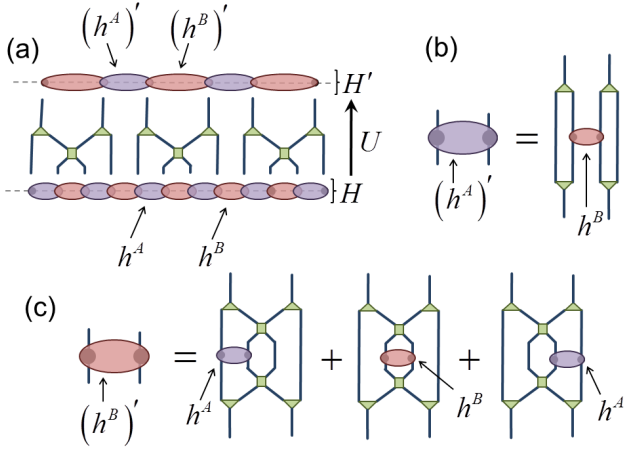}
\caption{(a) A single layer $U$ of a modified binary MERA, as depicted Fig. \ref{fig:DirectedMERA}(a), is used to coarse-grain the Hamiltonian $H$, as written in Eq. \ref{eq:AC1}, defined on the initial lattice $\mathcal L$ to a new Hamiltonian $H'$ defined on the coarser lattice $\mathcal L'$. (b) The coarse-grained coupling ${\left( {{h^A}} \right)^\prime }$ is obtained through the (disconnected) ascending superoperator $\mathcal A_D$, see also Eq. \ref{eq:AC3}. (c) The coarse-grained coupling ${\left( {{h^B}} \right)^\prime }$ is obtained through combination of the left, center and right ascending superoperators, $\mathcal A_L$, $\mathcal A_C$ and $\mathcal A_R$ respectively, see also Eq. \ref{eq:AC4}.}
\label{fig:ModScale}
\end{center}
\end{figure}

The scale invariant MERA offers a natural representation of the (scale-invariant) ground state of a gapless Hamiltonian at a critical point. Here we discuss the manifestation of scale-invariance in a specific MERA scheme, namely the modified binary MERA for one-dimensional systems (as introduced in Ref.\onlinecite{LongCFT1}), which is the one employed in this paper, see Fig. \ref{fig:DirectedMERA}(a). This scheme differs from previous implementations of the scale-invariant MERA \cite{Algorithms, Pfeifer} in several details. Most significantly, the coarse-graining scheme that the modified binary MERA arises from yields effective Hamiltonians that are translation invariant under shifts of two sites (even if the initial Hamiltonian was invariant under shifts of a single site), whereas previously considered schemes \cite{Algorithms, Pfeifer} yield effective Hamiltonians that are invariant under single site shifts. Consider a local $1D$ Hamiltonian $H$ of the form,
\begin{equation}
H = \sum\limits_{r{\text{ even}}} {h_{[r,r + 1]}^A}  + \sum\limits_{r{\text{ odd}}} {h_{[r,r + 1]}^B}, \label{eq:AC1}
\end{equation}
with $h^A$ and $h^B$ two potentially different nearest-neighbor couplings, and index $r$ labeling position on the lattice $\mathcal L$. Under coarse-graining with a single layer $U$ of the modified binary MERA, see Fig. \ref{fig:ModScale}(a), the Hamiltonian $H$ is mapped to a new Hamiltonian, $H \stackrel{U}{\longrightarrow} H'$, on a coarser lattice $\mathcal L'$, where the new Hamiltonian $H'$ is of the form,
\begin{equation}
H' = \sum\limits_{r{\text{ even}}} \left({h_{[r,r + 1]}^A}\right)'  + \sum\limits_{r{\text{ odd}}} \left({h_{[r,r + 1]}^B}\right)', \label{eq:AC2}
\end{equation}
for some new local couplings $\left({h^A}\right)'$ and $\left({h^B}\right)'$, and where index $r$ now labels position on $\mathcal L'$. The new coupling $\left({h^A}\right)'$ can be obtained through use of the (disconnected) ascending superoperator $\mathcal A_D$ on $h^B$,
\begin{equation}
{\left( {{h^A}} \right)^\prime } = {\mathcal A_D}\left( {{h^B}} \right), \label{eq:AC3}
\end{equation}
see Fig. \ref{fig:ModScale}(b), while the new coupling $\left({h^B}\right)'$ is obtained through use of (left, center, right) ascending superoperators $\mathcal A_L$ $\mathcal A_C$ and $\mathcal A_R$,
\begin{equation}
{\left( {{h^B}} \right)^\prime } = {A_L}\left( {{h^A}} \right) + {A_C}\left( {{h^B}} \right) + {A_R}\left( {{h^A}} \right), \label{eq:AC4}
\end{equation}
see Fig. \ref{fig:ModScale}(c). If the Hamiltonian $H$ is scale invariant fixed point of the MERA, and has had its energy spectrum shifted such that the ground state has zero energy, i.e. such that $\left\langle {{h^A}} \right\rangle  = \left\langle {{h^B}} \right\rangle  = 0$, then the couplings transform self-similarly under coarse-graining,
\begin{equation}
{\left( {{h^A}} \right)^\prime } = {h^A}/\Lambda,\; \; \; \; \; {\left( {{h^B}} \right)^\prime } = {h^B}/\Lambda, \label{eq:AC5}
\end{equation}
i.e. such that $H'=H/\Lambda$, where $\Lambda=2^z$ with $z$ is the dynamic critical exponent of $H$ (i.e. $z=1$ for a Lorentz invariant quantum critical point).

\begin{figure}[!tbph]
\begin{center}
\includegraphics[width=8.5cm]{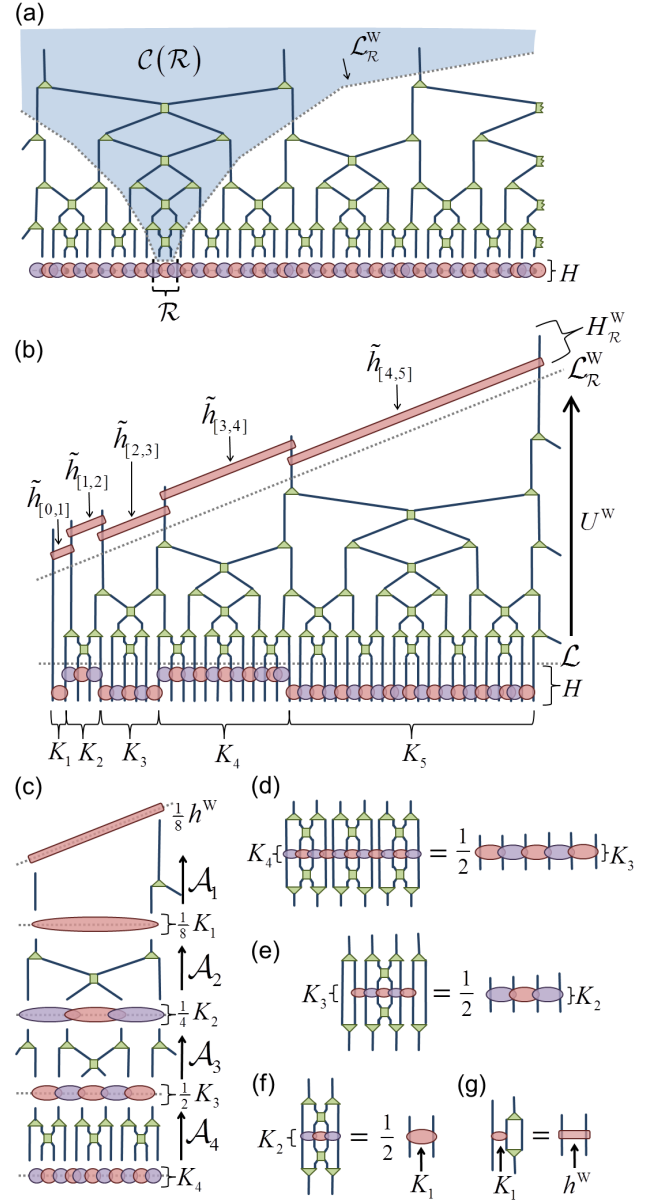}
\caption{(a) The causal cone $\mathcal C (\mathcal R)$ of a two-site region $\mathcal R \in \mathcal L$ in a modified binary MERA is shaded, and the associated Wilson chain $\mathcal L^{\Wilson}_{\mathcal R}$ is indicated. (b) The inhomogeneous coarse-graining $U^{\Wilson}$ maps the initial Hamiltonian $H$, here partitioned into shells $K_z$ of varying size (see Eqs. \ref{eq:AD1} and \ref{eq:AD2}), to the effective Hamiltonian $H^{\Wilson}_{\mathcal R}$ defined on the Wilson chain $\mathcal L^{\Wilson}_{\mathcal R}$. (c) A schematic depiction of the coarse-graining of a term from the local Hamiltonian $K_4$, assuming scale invariance of the Hamiltonian $H$, to a local coupling on the Wilson chain, see Eq. \ref{eq:AD6}. (d-f) Diagrammatic representations of the coarse-graining described in Eq. \ref{eq:AD7} for $s=4,3,2$. (g) A diagrammatic representation of $\mathcal A_1 (K_1) = h^{\Wilson}$.}
\label{fig:LogEffective}
\end{center}
\end{figure}

\section{Effective Hamiltonian for the Wilson chain} \label{ApdxD}
In the main text of this manuscript, it was asserted that the effective Hamiltonian $H_{\mathcal{R}}^{\Wilson}$ on the Wilson chain $\mathcal{L}^{\Wilson}_\mathcal{R}$, corresponding to the causal cone $\mathcal C(\mathcal R)$ of a local region $\mathcal R$ generically takes the form described in Eq. \ref{eq:hWilson} when the MERA describes a critical, scale-invariant state. Eq. \ref{eq:hWilson} is the same form as the effective one-dimensional Hamiltonian that Wilson obtained in his solution to the (three-dimensional) Kondo impurity problem. Here we derive Eq. \ref{eq:hWilson} explicitly by coarse-graining a one-dimensional Hamiltonian $H$ that is a scale-invariant fixed point of the MERA (in the modified binary scheme), and then outline how this derivation generalizes to systems in higher dimensions, see also Refs. \onlinecite{LongCFT2,BoundaryMERA} for complimentary derivations of the effective Hamiltonian for the Wilson chain.

For a modified binary MERA defined on a $1D$ lattice $\mathcal L$, we consider the Wilson chain $\mathcal L^{\Wilson}_{\mathcal R}$ associated to the two-site region $\mathcal R \in \mathcal L$ as shown in Fig. \ref{fig:LogEffective}(a). Let $\chi$ denote the dimension of each index connecting tensors in the MERA. Then the Wilson chain is a semi-infinite chain where each site has a vector space of dimension $\chi^2$, as two $\chi$-dimensional indices cross the surface of the causal cone between any depths $[s,s+1]$ in the MERA. The tensors in the MERA that are outside of the causal cone $\mathcal C(\mathcal R)$ implement a coarse-graining transformation $U^{\Wilson}$ that maps the initial Hamiltonian $H$ into the effective Hamiltonian $H^{\Wilson}_{\mathcal R}$ on the Wilson chain, see Fig. \ref{fig:LogEffective}(b). The effective Hamiltonian $H^{\Wilson}_{\mathcal R}$  can be written as,
\begin{equation}
H^{\Wilson}_{\mathcal R} = \sum\limits_{s = 0}^\infty  {\tilde h_{[s,s + 1]}(s)} \label{eq:AD4b}
\end{equation}
for some nearest-neighbor coupling $\tilde h(s)$ that depends explicitly on position $s$. We now examine how these local couplings $\tilde h(s)$ can be computed, and derive a relationship between couplings at different positions on the Wilson chain. For simplicity we shall consider only the contribution to the effective Hamiltonian $H^{\Wilson}_{\mathcal R}$ that comes from the half of $H$ that is to the right of the region $\mathcal R$ (noting the left half yields an identical contribution). Let us begin by rewriting the right half of $H$ as,
\begin{equation}
H^\textrm{right} = \sum\limits_{s = 1}^\infty  K_s, \label{eq:AD1}
\end{equation}
where $K_s$ denotes the sum of all terms in $H$ supported on the lattice $\mathcal L$ in the interval of sites of distance between $\left[ {r_s}, {r_{s + 1}}\right]$ to the right of $\mathcal R$, with $r_s$ defined as
\begin{equation}
{r_s} \equiv \left\{ {\begin{array}{*{20}{l}}
  {( {{2^{s + 1}} - 1} )/3,\ \ s\textrm{ odd}} \\
  {( {{2^{s + 1}} - 2} )/3,\ \ s\textrm{ even}} .
\end{array}} \right. \label{eq:AD2}
\end{equation}
For instance, $K_1$ is the sum of terms in the interval of sites at distance $\left[ {r_1}, {r_{2}}\right]=[1,2]$ from $\mathcal R$, which is just a single term,
\begin{equation}
K_1=h_{[1,2]}^B, \label{eq:AD3}
\end{equation}
while $K_2$ and $K_3$ are the sum of terms in the intervals of $\left[ {r_2}, {r_{3}}\right]=[2,5]$ and $\left[ {r_3}, {r_{4}}\right]=[5,10]$ respectively,
\begin{align}
K_2&=h_{[2,3]}^A + h_{[3,4]}^B + h_{[4,5]}^A,\nonumber \\
K_3&=h_{[5,6]}^B + h_{[6,7]}^A + h_{[7,8]}^B + h_{[8,9]}^A + h_{[9,10]}^B, \label{eq:AD4}
\end{align}
and so forth, see also Fig. \ref{fig:LogEffective}(b). Let $\mathcal A_s$ denote the ascending superoperator that implements coarse-graining of the Hamiltonian term $K_s$ through one layer of the MERA (the explicit forms of $\mathcal A_4$, $\mathcal A_3$, $\mathcal A_2$ and $\mathcal A_1$ are depicted in Fig. \ref{fig:LogEffective}(d-g)). Then the local coupling $\tilde h_{[s,s+1]} (s)$ of the effective Hamiltonian is obtained by coarse-graining $K_{s+1}$ a total of $s+1$ times,
\begin{equation}
\tilde h_{[s,s+1]} (s) = \left( \mathcal A_1 \circ \mathcal A_2 \circ \cdots \circ \mathcal A_{s} \circ \mathcal A_{s+1} \right) \left( K_{s+1} \right). \label{eq:AD5}
\end{equation}
In Fig. \ref{fig:LogEffective}(c) we depict the coarse-graining of the term $K_4$, a particular case of Eq. \ref{eq:AD5}, which is written as,
\begin{equation}
\tilde h_{[3,4]} (3) = \left( \mathcal A_1 \circ \mathcal A_2 \circ \mathcal A_3 \circ \mathcal A_4 \right) \left( K_4 \right). \label{eq:AD6}
\end{equation}
If the local Hamiltonian $H$ is invariant under coarse-graining with the MERA, as discussed in the previous section (see, in particular, Eq. \ref{eq:AC5}), then it can be seen that the $K$ terms transform in a precise way under coarse-graining,
\begin{equation}
\mathcal A_{s+1} \left( K_{s+1} \right) = \tfrac{1}{2^{z}} K_{s}, \label{eq:AD7}
\end{equation}
for all $s \ge 1$. Here $z$ is the dynamic critical exponent of $H$. If we define $h^{\Wilson} \equiv \mathcal A_1 \left( K_{1}\right)$ then all local couplings $\tilde h_{[s,s+1]}(s)$ of the effective Hamiltonian $H^{\Wilson}_{\mathcal R}$, as written in Eq. \ref{eq:AD4b}, are all proportionate to this $h^{\Wilson}$,
\begin{equation}
\tilde h_{[s,s+1]}(s) = \frac{1}{2^{zs}} h^{\Wilson}_{[s,s+1]}. \label{eq:AD8}
\end{equation}
Thus the effective Hamiltonian $H^{\Wilson}_{\mathcal R}$ for the Wilson chain is consistent with that proposed in Eq. \ref{eq:hWilson}, with the geometric decay of coupling strength $\Lambda=2^z$.

The essential features of the above derivation, which are geometric in nature, hold for MERA defined on higher dimensional lattices such that they also yield effective Hamiltonians of the form of Eq. \ref{eq:hWilson} on their corresponding Wilson chains, as we now outline. Let us assume that we have a $d$-dimensional hypercubic lattice $\mathcal L$ on which a local Hamiltonian $H=\sum h$ and a scale invariant MERA are defined, and that we would like to understand the effective Hamiltonian $H^{\Wilson}_{\mathcal R}$ for the Wilson chain ${\mathcal L}^{\Wilson}_{\mathcal R}$ associated to a local region $\mathcal R$.

\begin{figure}[!tbph]
\begin{center}
\includegraphics[width=8.5cm]{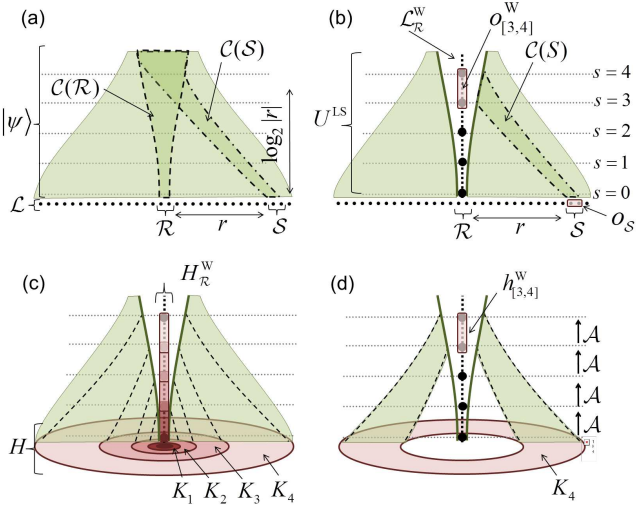}
\caption{(a) Given a MERA $\ket{\psi}$ defined on lattice $\mathcal L$, the causal cones $\mathcal C (\mathcal R)$ and $\mathcal C (\mathcal S)$ of two local regions $\mathcal R$ and $\mathcal S$, displaced from each other by some distance $r$, intersect and become coincident after depth $s\approx \log_2 \left| r \right|$. (b) A local operator $o_{\mathcal S}$ supported on $\mathcal S$ is coarse-grained under $U^{\Wilson}$ to a new operator $o^{\Wilson}$ supported on the region of the Wilson chain $\mathcal L^{\Wilson}_{\mathcal R}$ where $\mathcal C (\mathcal S)$ intersected $\mathcal C (\mathcal R)$. (c) A local Hamiltonian $H$, here partitioned into pieces $K_z$ supported on concentric shells about $\mathcal R$, is mapped to a coarse-grained Hamiltonian $H^{\Wilson}$ on the Wilson chain. (d) The piece $K_4$ of $H$, consisting of the sum of terms in $H$ that are supported the shell between $r=2^{3}$ and $r=2^4$ sites distant from $\mathcal R$, is coarse-grained into a two-body coupling $h^{\Wilson}_{[3,4]}$ on the Wilson chain $\mathcal L^{\Wilson}_{\mathcal R}$, see also Eq. \ref{eq:InhomoCG}.}
\label{fig:Mapping}
\end{center}
\end{figure}

Given a local region $\mathcal S$ in $\mathcal L$ that is displaced by vector $\vec r$ from $\mathcal R$, the causal cones of the two regions will intersect roughly at depth $\sim \log_2 \left| \vec r \right|$ (note that we are assuming that each layer of the MERA rescales the lattice by a factor $\tfrac{1}{2}$ in all spatial dimensions, as with the modified binary MERA), see Fig. \ref{fig:Mapping}(a-b). The depth at which the causal cones intersect informs us the scale at which an operator $o_{\mathcal S}$ that is supported on $\mathcal S$ is coarse-grained onto the Wilson chain ${\mathcal L}^{\Wilson}_{\mathcal R}$ associated to region $\mathcal R$. Thus one can partition the lattice $\mathcal L$ into a series of concentric (hypercubic) shells $\mathcal R_s$ about the local region $\mathcal R_0 \equiv \mathcal R$, where shell $\mathcal R_s$ is roughly comprised of all sites at a distance between $2^{s-1}$ and $2^s$ sites away from $\mathcal R_0$, such that any operator that is supported on the shell $\mathcal R_s$ will be coarse-grained to a new local operator supported on sites $[s-1,s]$ of the Wilson chain ${\mathcal L}^{\Wilson}_{\mathcal R}$. Let us, as with Eq. \ref{eq:AD1} for the $1D$ MERA, rewrite the initial Hamiltonian as $H = \sum\nolimits_{s=0}^\infty K_s$ where each $K_s$ corresponds to the sum of all the couplings supported on the shell $\mathcal R_s$, see Fig. \ref{fig:Mapping}(c). It is then seen that the coupling $\tilde h_{[s,s+1]} (s)$ in the effective Hamiltonian for the Wilson chain arises through $s+1$ coarse-grainings of each $K_{s+1}$, i.e.
\begin{equation}
\tilde h_{[s,s+1]} (s) = \left( \mathcal A_1 \circ \mathcal A_2 \circ \cdots \circ \mathcal A_{s} \circ \mathcal A_{s+1} \right) \left( K_{s+1} \right). \label{eq:InhomoCG}
\end{equation}
where each $\mathcal A_s$ represents the appropriate ascending superoperator that coarse-grains $K_s$ through one layer of the MERA, see Fig. \ref{fig:Mapping}(d). Roughly speaking, the term $K_{s+1}$ collects together $O(2^{d(s+1)})$ nearest neighbor terms in $H$, each of which has is then coarse-grained $(s+1)$ times to give the effective coupling of the Wilson chain $\tilde h_{[s,s+1]} (s)$. In a critical system in $d$ space dimensions the scaling dimension of a single Hamiltonian term is $\Delta = d+z$, where $z$ is the dynamic critical exponent of $H$, with $z=1$ for Lorentz invariant quantum critical points. This implies that one such term is reduced by a factor $2^{-\Delta} = 2^{-(d+z)}$ with each coarse-graining step. Hence the effective couplings are of the form,
\begin{equation}
\tilde h_{[s,s+1]}(s) = \Lambda^{-s} h^{\Wilson}_{[s,s+1]}. \label{eq:AD8b}
\end{equation}
where the independence of $h^{\Wilson}_{[s,s+1]}$ on $s$ follows from the invariance of $H$ both under translations and re-scaling transformations, the amplitude $\Lambda^{-s}$ results from,
\begin{equation}\label{eq:Lambda}
    O(2^{ds}) \times \left(\frac{1}{2^{d+z}}\right)^{s} \approx 2^{-sz} = \Lambda^{-s}.
\end{equation}
Notice that Eq. \ref{eq:AD8b} is the same as Eq. \ref{eq:AD8}, which was derived explicitly for a one-dimensional system.

\begin{figure}[!tbph]
\begin{center}
\includegraphics[width=8.5cm]{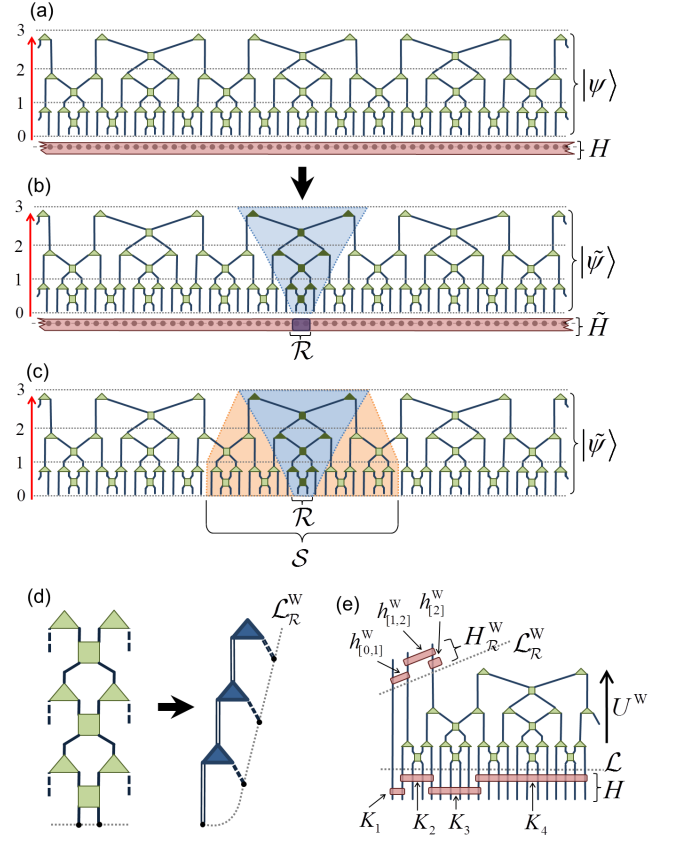}
\caption{(a) The ground state of the gapped Hamiltonian $H$ is represented by a finite correlation length MERA $\ket{\psi}$, here with $s_\textrm{max}=3$ layers. (b) Directed influence asserts that the ground state MERA $\ket{\tilde \psi}$ for the Hamiltonian modified with a local impurity $\tilde H\equiv H +H_{\mathcal R}^{\textrm{imp}}$ can possess the same tensors as $\ket{\psi}$ everywhere outside of the causal cone $\mathcal C(\mathcal R)$ of the local region $\mathcal R$. (c) Changing the tensors within only within a causal cone $\mathcal C(\mathcal R)$ of the MERA only affects the expectation values of local observables that are supported within the larger region $\mathcal S$, see also Eq. \ref{eq:AE2}. (d) The set of tensors within the causal cone of $\mathcal R$ are equivalent to an MPS on the Wilson chain $\mathcal L^{\Wilson}_{\mathcal R}$, here a finite $1D$ lattice of four sites. (e) The inhomogeneous coarse-graining $U^{\Wilson}$ maps the initial Hamiltonian $H$, here partitioned into shells $K_s$ of varying size (see Eqs. \ref{eq:AD1} and \ref{eq:AD2}), to the effective Hamiltonian $H^{\Wilson}$ for the Wilson chain. Notice that terms $K_s$ for $s\ge 4$ contribute only irrelevant additive constants to the effective Hamiltonian.}
\label{fig:LogGapped}
\end{center}
\end{figure}

\section{Minimal updates in a finitely correlated MERA} \label{ApdxE}
In the main text we have discussed a theory of minimal updates in the holographic description of a many-body ground state. For simplicity, we have considered a translation invariant system that is gapless fixed-point of the RG flow, and therefore invariant under changes of scale (as implemented by means of discrete coarse-graining transformations). However, the essential parts of our arguments do not rely on translation or scale invariance, and the proposed minimal updates also apply in the absence of such space symmetries.

In this appendix we address the case of a gapped system in a topologically trivial phase, where the ground state can be described by a \textit{finitely correlated} MERA \cite{Algorithms}. A finitely correlated MERA has a set number of layers $s_\textrm{max}$, see Fig. \ref{fig:LogGapped}(a), and is expected to offer a good approximation to the ground state of a gapped Hamiltonians $H$ (in a topologically trivial phase) when the correlation length $\xi$ fulfills $\xi < 2^{s_\textrm{max}}$.

The justification for directed influence in finitely correlated MERA and its consequence in permitting a minimal update of the MERA under a local change to the Hamiltonian is analogous to the case analyzed in the main text. However, some implications of directed influence are different. One difference is that modifying a finitely correlated MERA within a causal cone $\mathcal C(\mathcal R)$ only affects the ground state properties within some localized region around $\mathcal R$. Consider, for instance, taking the expected value of a local observable $o$ from two different finitely correlated MERA $\ket{\psi}$ and $\ket{\tilde \psi}$, each with a fixed number $s_\textrm{max}$ layers, whose tensors only differ within the causal cone $\mathcal C(\mathcal R)$ of a local region $\mathcal R \in \mathcal L$. Here, one can identify a larger region $\mathcal S\in \mathcal L$, which is defined as the set of all sites whose causal cone intersects with $\mathcal C (\mathcal R)$ (notice that this roughly corresponds to the shell of thickness $\sim 2^{s_\textrm{max}}$ about $\mathcal R$, see Fig. \ref{fig:LogGapped}(c)), such that the expectation value of any local observable is identical between the two MERA whenever the observable is outside the support of $\mathcal S$, i.e.
\begin{equation}
\big \langle \psi \big| {o_{[r]}}\big| \psi  \big\rangle  = \big\langle {\tilde \psi } \big|{o_{[r]}}\big| {\tilde \psi } \big\rangle ,\ \ \ \ r \notin \mathcal{S} \label{eq:AE2}
\end{equation}
for all local observables $o$. Recall that, in the case of scale invariant MERA, changing the tensors within a causal cone can affect the expectation value of a local observable everywhere on the lattice $\mathcal L$. This difference arises as it is only in finitely correlated MERA that separated regions of the lattice can be causally disconnected, i.e. such that there is no overlap in the respective causal cones of the regions.

Another difference in dealing with a finitely correlated MERA is that their corresponding Wilson chains are finite $1D$ lattices of $(s_\textrm{max}+1)$ sites, see Fig. \ref{fig:LogGapped}(d)), as opposed to the semi-infinite $1D$ lattices that arise from scale invariant MERA. Let us examine, given a local Hamiltonian $H$ defined on the lattice $\mathcal L$, the computation of the effective Hamiltonian $H^{\Wilson}_{\mathcal R}$ corresponding to a local region $\mathcal R \in \mathcal L$. It can be seen that only part of the local Hamiltonian $H$ near the region $\mathcal R$ contributes to the effective Hamiltonian; specifically, if we once more partition the local Hamiltonian into terms $K_s$ supported on a series of concentric shells, as described by Eqs. \ref{eq:AD1} and \ref{eq:AD2}, then it is only terms $K_s$ for $s\le s_\textrm{max}$ that are coarse-grained into couplings on the effective Hamiltonian $H^{\Wilson}_{\mathcal R}$ (whereas terms $K_s$ for $s>s_\textrm{max}$ only shift the overall energy levels of $H^{\Wilson}_{\mathcal R}$ by an irrelevant constant, see Fig. \ref{fig:LogGapped}(e)).


\begin{thebibliography}{99}

\bibitem{RGA}
K.G. Wilson, Adv. Math., Volume 16, Issue 2, Pages 170-186 (1975).
\bibitem{RGB}
K.G. Wilson, Rev. Mod. Phys. 47, 773 (1975).
\bibitem{RGC}
M.E. Fischer, Rev. Mod. Phys. 70, 653 (1998).

\bibitem{AdSCFTA}
J. Maldacena, Adv. Theor. Math. Phys. 2, 231 (1998).
\bibitem{AdSCFTB}
E. Witten, Adv. Theor. Math. Phys. 2, 253 (1998).
\bibitem{AdSCFTC}
R. Bousso, Rev. Mod. Phys. 74, 825 (2002).

\bibitem{ERA}
G. Vidal, Phys. Rev. Lett. 99, 220405 (2007).
\bibitem{ERB}
For an introduction, see G. Vidal, "Entanglement Renormalization: an introduction", chapter 5 in the book "Understanding Quantum Phase Transitions", edited by Lincoln D. Carr (Taylor $\&$ Francis, Boca Raton, 2010).

\bibitem{MERA}
G. Vidal, Phys. Rev. Lett., 101, 110501 (2008).

\bibitem{WhatIsHolography}
By holographic description we mean one that extends in an additional dimension corresponding to scale or RG flow. The MERA is then a holographic description in this general sense. The term 'holography' may also be used in a more restricted sense, where one expects the scale dimension to be locally equivalent to the other space dimensions in the bulk. By studying properties of the MERA, we expect to learn about the general structure of RG-based holographic descriptions. Examples of those are modularity \cite{LongCFT2}, and the mapping of a global symmetry at the boundary into a local symmetry in the bulk \cite{Singh}.

\bibitem{Swingle}
B. Swingle, Phys. Rev. D 86, 065007 (2012).

\bibitem{BranchingA}
G. Evenbly, G. Vidal, Phys. Rev. Lett. 112, 220502 (2014).
\bibitem{BranchingB}
G. Evenbly, G. Vidal, Phys. Rev. Lett. 112, 240502 (2014).

\bibitem{Swingle2}
B. Swingle, pre-print arXiv:1209.3304.

\bibitem{Metric}
M. Nozaki, S. Ryu, and T. Takayanagi, JHEP 10, 193 (2012).

\bibitem{AllA}
G. Evenbly, G. Vidal, J. Stat. Phys. 145, 891 (2011).
\bibitem{AllB} 
J. Molina-Vilaplana and P. Sodano, JHEP 10, 11 (2011).
\bibitem{AllC}
C. Beny, New J. Phys. 15 (2013) 023020.
\bibitem{AllD} 
H. Matsueda, M. Ishihara, and Y. Hashizume, Phys. Rev. D 87, 066002 (2013).
\bibitem{AllE}
T. Hartman, J. Maldacena, JHEP 5, 14 (2013).

\bibitem{Singh}
S. Singh, G. Vidal, Phys. Rev. B 88, 121108(R) (2013).

\bibitem{LongCFT2}
G. Evenbly, G. Vidal, J. Stat. Phys. 157, 931 (2014).

\bibitem{dVersusdPlus1}
In the Hamiltonian formalism, a $d$-dimensional lattice describes a system in $d+1$ space-time dimensions. Thus, a critical ground state $\ket{\psi}$ in $d$ space dimensions and its MERA representation in $d+1$ dimensions (in both cases, there is no time dimension, given that the time evolution of a ground state is trivial), should be thought of as analogous to a CFT in $d+2$ dimensions and AdS in $d+2$ dimensions, respectively.

\bibitem{NRGreview}
R. Bulla, T. Costi, and T. Pruschke, Rev. Mod. Phys. 80, 395 (2008).

\bibitem{ManyMERASchemes}
There are many possible MERA schemes even in just $d=1$ dimensions, depending on aspects such as how many sites are coarse-grained into a single site, or where the disentanglers are placed, see Ref. \onlinecite{LongCFT1}.

\bibitem{Giovannetti}
V. Giovannetti, S. Montangero, R. Fazio, Phys. Rev. Lett. 101, 180503 (2008).

\bibitem{Pfeifer}
R.N.C. Pfeifer, G. Evenbly, and G. Vidal, Phys. Rev. A 79, 040301(R) (2009).

\bibitem{LongCFT1}
G. Evenbly and G. Vidal, "Quantum Criticality with the Multi-scale Entanglement Renormalization Ansatz", chapter 4 in the book "Strongly Correlated Systems. Numerical Methods", edited by A. Avella and F. Mancini (Springer Series in Solid-State Sciences, Vol. 176 2013), arXiv:1109.5334.

\bibitem{Algorithms}
G. Evenbly and G. Vidal, Phys. Rev. B 79, 144108 (2009). 

\bibitem{ModularityAlsoForLargeR}
The minimal updates discussed in this paper are not restricted to a small region $\mathcal{R}$. For instance, in boundary or interface problems, region $R$ can correspond to a semi-infinite line, plane, etc \cite{LongCFT2}.

\bibitem{ConformalDefect} M. Oshikawa and I. Affleck, Nucl. Phys. B 495:533-582 (1997).

\bibitem{BoundaryMERA}
G. Evenbly, R. N. C. Pfeifer, V. Pico, S. Iblisdir, L. Tagliacozzo, I. P. McCulloch, G. Vidal, Phys. Rev. B 82, 161107(R) (2010).

\bibitem{MPSA}
M. Fannes, B. Nachtergaele, and R. F. Werner, Commun. Math. Phys. 144, 443 (1992).
\bibitem{MPSB}
S. Ostlund and S. Rommer, Phys. Rev. Lett. 75, 3537 (1995).

\bibitem{DMRG}
S.R. White, Phys. Rev. Lett. 69, 2863 (1992).

\bibitem{NoDMRG}
If the update were not restricted to the causal cone then a variational algorithm based upon optimally preserving the support of the ground state's reduced density matrix \cite{DMRG} would likely yield location-dependent tensors $(\tilde{u}(x), \tilde{w}(x))$, given that the reduced density matrix $\tilde{\rho}_{\mathcal{S}}$ of a region $\mathcal{S}$ depends on the distance between $\mathcal{S}$ and the location $\mathcal{R}$ of the impurity. Thus the Hamiltonian $\tilde{H}=H + H_{\mathcal{R}}^{\imp}$ will be coarse-grained into an effective Hamiltonian made of location-dependent terms: the initially localized impurity has effectively become delocalized. That the location-independent bulk tensors $(u, w)$ already do a good (though perhaps not optimal) job of preserving the support $\tilde{\rho}_{\mathcal{S}}$ for all $\mathcal{S}$ away from the impurity is a key result of \emph{minimal updates}.

\end{thebibliography}
\end{document}